\documentclass[aps,prl,twocolumn,preprintnumbers,amsmath,showpacs,amssymb]{revtex4}
\usepackage{graphicx}
\usepackage{dcolumn}
\usepackage{bm}

\begin{document}
\vspace{-0.3in}
\title{Integer quantum Hall effect on a six valley
hydrogen-passivated silicon (111) surface}

\author{K. Eng}
\author{R. N. McFarland}%
\author{B. E. Kane}
\affiliation{Laboratory for Physical Sciences, University of Maryland
at College Park,College Park, MD 20740 USA}%

\date{\today}
\begin{abstract}

We report magneto-transport studies of a two-dimensional electron
system formed in an inversion layer at the interface between a
hydrogen-passivated Si(111) surface and vacuum. Measurements in
the integer quantum Hall regime demonstrate the expected sixfold
valley degeneracy for these surfaces is broken, resulting in an
unequal occupation of the six valleys and anisotropy in the
resistance. We hypothesize the misorientation of Si surface breaks
the valley states into three unequally spaced pairs, but the
observation of odd filling factors, is difficult to reconcile with
non-interacting electron theory.

\end{abstract}

\maketitle
   The silicon field effect transistors (FETs) that are at the heart
of contemporary microelectronics rely on mobile electrons or holes
confined at the interface between Si and a higher bandgap barrier
material.  This barrier in metal oxide silicon (MOS) FETs is
SiO$_2$, an amorphous material which introduces inevitable
disorder at the Si-SiO$_2$ interface and limits the carrier
mobility in these devices.  A crystalline interface can be created
using epitaxial SiGe-Si layers in which mobilities can be over an
order of magnitude higher than the best MOSFET
devices\cite{schaffler}, but this technique is limited to the
[100] oriented surfaces\cite{lee}. Recently a new technique for Si
crystalline interfaces has been demonstrated\cite{eng1} in which a
Si interface is passivated with a monolayer of hydrogen and the
barrier material is a vacuum. While the inertness and high degree
of atomic perfection of these surfaces has been known for some
time\cite{higashi,yabl,hinesrev}, the development of high mobility
electronic devices on H-Si enables the exploration of
two-dimensional (2D) physics of novel Si surface orientations and
may one day allow quantum devices to be engineered at the atomic
scale using surface manipulation techniques\cite{qubit}.

    We report here the first detailed magneto-transport studies of
a 2D electron system (2DES) at a H-Si(111) surface gated through a
vacuum barrier. Electron mobilities are an order of magnitude
higher ($24,000$cm$^2$/Vs) than Si(111) MOSFETs, enabling the
observations of the integer quantum Hall effect (IQHE). In the
effective mass approximation the ground state for a 2DES on the
Si(111) surface is six-fold degenerate with each Si conduction
band valley contributing an equal number of carriers, (Fig. 1d)
each with anisotropic in-plane masses, $m_{x} = 0.19m_{o}$ and
$m_{y} = 0.67m_{o}$\cite{ando} ($m_o$ is the mass of the free
electron). However, measurements of Shubnikov-de Haas (SdH)
oscillations in Si(111) MOSFETs have shown conflicting valley
degeneracies of two\cite{tsui,dorda,ando} and six\cite{tsui2,cole}
along with isotropic resistivities for both. Subsequent
proposals\cite{tsui2,kelly} have tried to explain these anomalous
observations, but to date conclusive experimental results are
still lacking.

\begin{figure}[t]
\begin{center}
  \includegraphics[height=3.15in,width=3.1in]{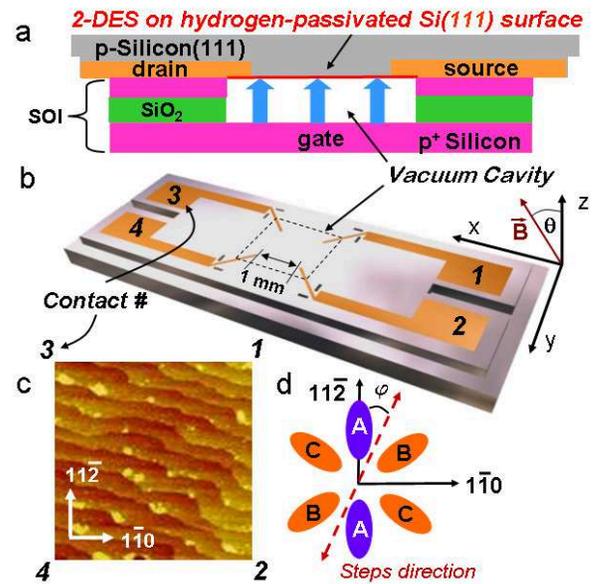}
  \caption{(a) Schematic cross-section of a
 H-Si(111) substrate contact bonded to a SOI substrate.
A p$^+$ layer in the SOI defines the gate, where blue arrows
depict the electric field.  A 2DES is formed at the H-Si(111)
surface within an encapsulated cavity. (b) The H-Si(111) substrate
has four n$^+$ contacts numbered accordingly. Tilted magnetic
fields are applied in the $x$-$z$ plane. (c) A 1 $\mu$m $\times$ 1
$\mu$m AFM image of atomic steps on a H-Si(111) surface in
relation to the crystal directions and the contacts of the device.
(d) The projection of the six valleys for the Si(111) surface with
pairs of valleys labeled A, B and C. } \label{fig1}
\end{center}
\vspace{-28pt}
\end{figure}

The high mobility 2DES can be created by contact bonding two
individual Si substrates\cite{eng1} (Fig. 1a). One is the
H-Si(111) substrate (float zone, p-type, $\rho \sim 10$
$\Omega$-cm) which has four phosphorous contacts forming a
1-mm-wide square with sides oriented parallel to the [$1\bar{1}0$]
and [$11\bar{2}$] crystallographic directions (Fig. 1b and 1c).
The second is a silicon-on-insulator (SOI) substrate which acts as
the remote gate, where an electric field can be controlled within
an etched cavity. The Si(111) surface is H-passivated by immersion
in an ammonium fluoride solution. The two substrates are then
bonded in vacuum ($\sim$10$^{-6}$ Torr), which allows the remote
gate to induce electrons on the H-Si(111) surface and protects the
air sensitive surface inside the cavity.



    Figure 2a shows traces of the Hall ($R_{xy}$) and the
longitudinal resistance ($R_{xx}$) as a function of perpendicular
magnetic field, $B_{\perp}$, for a constant electron density, $n =
6.5\times10^{11}$ cm$^{-2}$ at $T =300$ mK. We have obtained
similar data in two devices over a wide range of $n$. We present
data for a single device at a density where IQHE features are most
apparent. There are two orientations in which $R_{xx}$ is
measured; $R_{(12,34)}$ and $R_{(13,24)}$, where the first and
second subscripts represent the current and voltage contacts
respectively (Fig. 1b). Anisotropy in $R_{xx}$ is observed, but
both orientations exhibit minima that occur at integer values of
filling factor, $\nu = nh/eB_{\perp}$. Similar behavior (within
3\%) is also observed for R$_{(43,21)}$ and R$_{(42,31)}$. These
minima are directly correlated with the observed plateaus in
$R_{xy}$, a trademark of the QHE. Filling factors less than six
are observed at $\nu$ = 5, 4, 3, and 2, where $\nu = 5$ is
apparent through $d^{2}R_{xx}/dB_{\perp}^{2}$. Figure 2b shows
characteristic behavior at low $B$ fields: $R_{xy}/B_{\perp}$ is
dependent on $B_{\perp}$; both $R_{(12,34)}$ and $R_{(13,24)}$
display a positive magneto-resistance (MR); finally, while more
apparent in $R_{(12,34)}$, both orientations display SdH
oscillations with minima occurring approximately every eighth
Landau level (LL) filling factor, $\Delta \nu \sim 8$.

\begin{figure}[!]
\begin{center}
  \includegraphics[height=4.5in,width=3.1in,clip]{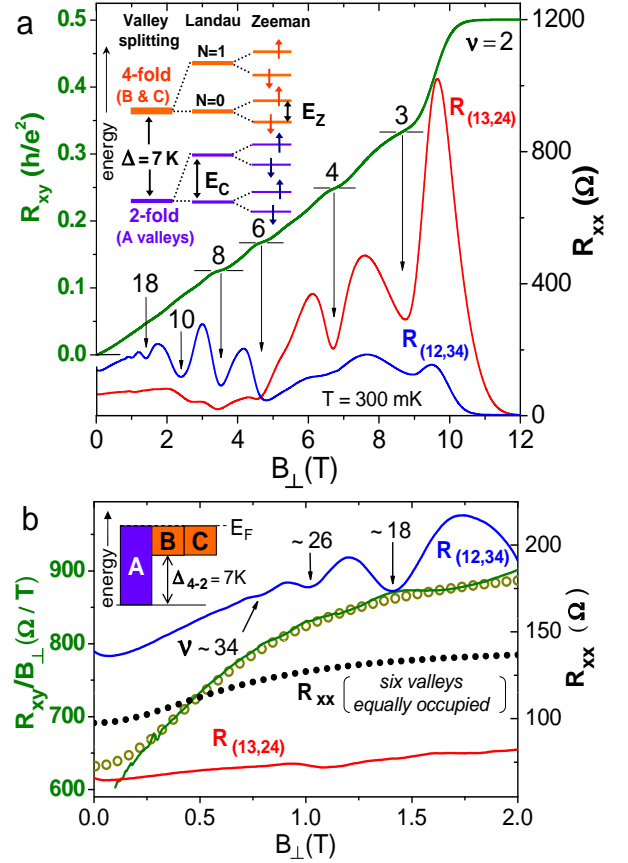}
  \caption{(a) $R_{xy}$ (green), $R_{(12,34)}$ (blue) and $R_{(13,24)}$ (red)
versus $B_{\perp}$ for the device at $T = 300$ mK and $n = 6.5
\times 10^{11}$ cm$^{-2}$. Plateaus in $R_{xy}$ are labeled by
horizontal bars and the numbers indicate their corresponding
filling factor $\nu$. The inset is a schematic energy level
diagram depicting the three energy scales affecting
$\mathcal{D}(E)$ at $B_{\perp} = 1$ T. (b) The same data from (a)
is plotted within $0\leq B_{\bot}\leq 2$ T. Hall resistance is
plotted as $R_{xy}/B_{\perp}$ and the divergence is removed for
$B_{\perp} \leq0.1$ T. Open symbols ($R_{xy}/B_{\perp}$) and
closed symbols ($R_{xx}$) correspond to calculations having all
six valleys equally occupied. The inset depicts $\mathcal{D}(E)$
for the 7 K model at $B = 0$. } \label{fig2}
\end{center}
\vspace{-26pt}
\end{figure}

    Classical MR calculations\cite{smith} show that in
a multi-component electron system in which electrons have
differing anisotropic in-plane masses, $R_{xy}$ is non-linear with
respect to $B_{\perp}$, and $R_{xx}$ should display a positive MR.
In Figure 2b we plot $R_{xy}/B_{\perp}$ and $R_{xx}(B)$ overlaid
with a classical MR model in which all six valleys are equally
occupied, with $n$ and the mean scattering time, $\tau$, as free
parameters. The values which provided the best fit to $R_{xy}$ are
within 5\% of observed values: $\tau = 3.5$ ps and $n \simeq  6.8
\times 10^{11}$ cm$^{-2}$.  This model agrees well with the
observed behavior in $R_{xy}$ and the positive MR in $R_{xx}$.
However, in order to explain the observed anisotropy in $R_{xx}$,
an unequal population of electrons among the six valleys is
required.

\begin{table}[!]
\begin{center}
\begin{tabular}{c c c c c}
\hline \hline
  \ $\nu$ \ & \ n ($\times10^{11}$ cm$^{-2}$) \ & \ $\theta$ \ & \ $\Delta_{(13,24)}$(K)\  & \ $\Delta_{(12,34)}$(K)\ \\
  \hline
  2 & 5.69 &\ 0$^{\circ}$   \ & $8\pm0.5$ & $7\pm0.5$ \\
  3 & 6.50 &\ 39$^{\circ}$  \ & $0.4\pm0.2$ & $0.8\pm0.2$ \\
  4 & 6.50 &\ 48.4$^{\circ}$\ & $1.1\pm0.2$ & $1.2\pm0.2$ \\
  6 & 6.50 &\ 48.4$^{\circ}$\ & $1.1\pm0.2$ & $1.2\pm0.2$ \\
  \hline \hline
\end{tabular}
\caption{Energy gaps for various filling factors, where $\theta
=0^{\circ}$ corresponds to a magnetic field normal to the 2DES.
The subscripts in $\Delta_{(12,34)}$ represent the measurement
orientation.} \label{activation}
\end{center}
\vspace{-26pt}
\end{table}

    The strong $R_{xx}$ minimum at $\nu = 2$, suggests a
gap between the six valleys.  We obtain gap energies,
$\Delta_{\nu}$, from the temperature dependence of $R_{xx}$,
$R_{xx} \sim$ exp$(-\Delta_{\nu}/2k_{B}T)$.  These measurements
show that the energy gap, $\Delta_{2}$, is insensitive to small
changes in $B_{\perp}$ and in-plane fields, $B_{\parallel}$ (Table
1). It is by far the largest gap ($\simeq 7$ K) and provides
strong evidence that the six valleys are indeed broken
asymmetrically following a 4-2 valley configuration.  For
simplicity we will assume a model which is composed of only one
type of splitting which is independent of $B$: two valleys (A) are
7 K lower in energy than the remaining four valleys (B and C) in
the ground state (inset of Fig. 2b). The inset of Fig. 2a
illustrates the effect $B_{\perp}$ has on the the density of
states, $\mathcal{D}(E)$, for this `7 K model'. Quantization of
the electron's orbital motion creates LLs which are separated by
the cyclotron gap, $E_C = \hbar e B_{\perp}/m_{\mathcal{D}}$,
where $m_{\mathcal{D}} = (m_{x}m_{y})^{1/2} = 0.358m_o$ is the
density of states mass for Si(111). Each LL then splits into two
energy levels due to spin. This energy separation is the Zeeman
energy, $E_Z = g^{*}\mu_{B}B_{\perp}/\cos \theta$, where $g^*$ is
the effective g-factor and $\mu_B$ is the Bohr magneton.

\begin{figure}[!]
\begin{center}
  \includegraphics[height=4.75in,width=3.1in]{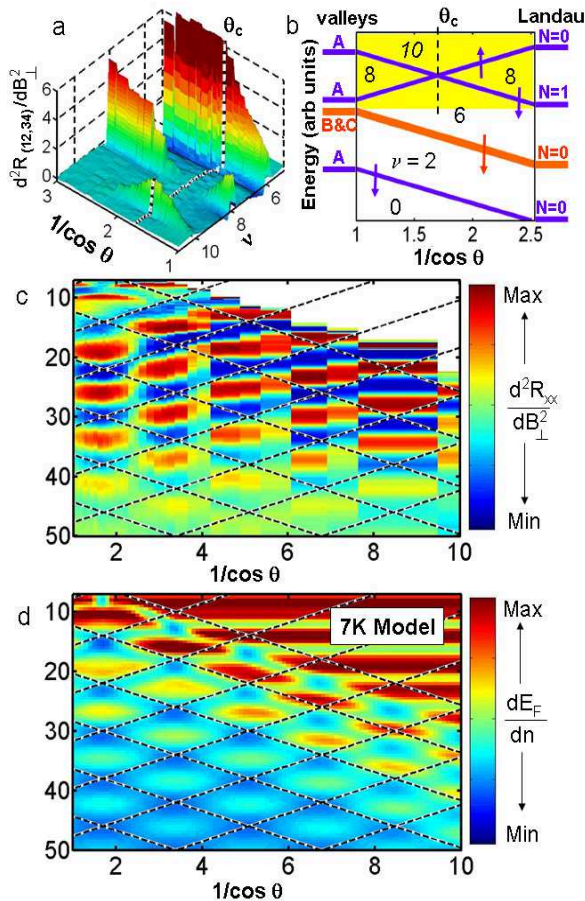}
\caption{Data and calculations represented at $n =
6.5\times10^{11}$cm$^{-2}$ and $T = 300$ mK. For the color scales
represented in (a), (c) and, (d), a maximum in
$d^{2}R_{xx}/dB_{\perp}^2$ or $dE_F/dn$ corresponds to a minimum
in $R_{xx}$.  (a) A surface plot of $d^2R_{(12,34)}/d\nu^2$ versus
$1/\cos\theta$ for filling factors within $5\leq\nu\leq 11$.  At
$\theta_c$, the $R_{xx}$ minimum at $\nu=8$ disappears while a
minimum at $\nu=10$ is at a peak. (b) Schematic energy level
diagram following the 7 K model predicting the two-fold energy
levels from the A valleys (spin up and spin down) crossing one
another at $\theta_c$. (c) A plot of $d^2R_{(12,34)}/dB_{\perp}^2$
versus $\nu$ and $1/\cos \theta$. The overlaid schematic fan
diagram (dashed lines) represents expectations from the 7 K model
with $\Delta\nu\simeq8$. (d) A plot of a calculated $dE_{F}/dn$
versus $\nu$ and $1/\cos \theta$ using the 7 K model. The same fan
diagram used in (c) is overlaid on top of calculations. }
\label{fig3}
\end{center}
\vspace{-26pt}
\end{figure}

    The validity of this 7 K model can be tested
with measurements in $R_{xx}$ versus tilted $B$ fields.  Rotating
the device some angle, $\theta$, away from $B_{\perp}$ changes
$E_Z$ with respect to $E_C$, and eventually energy levels of
different spins will overlap with one another. The angles at which
these coincidences occur can be observed through the disappearance
of minima in $R_{xx}$. One such coincidence is shown in Fig. 3a,
where a minimum in $d^2R_{(12,34)}/dB_{\perp}^2$ corresponds to
the disappearance of the $R_{xx}$ minimum at $\nu = 8$ and a
concomitant maximum occurs at $\nu=10$.  Similar behavior is also
observed for $d^{2}R_{(13,24)}/dB_{\perp}^2$.  The second
derivatives of $R_{xx}$ are plotted in order to amplify structure
and to eliminate the contribution from the positive MR at low $B$
fields. Figure 3b depicts a schematic energy level diagram
following the 7 K model and using $B_{\perp}$ values centered at
$\nu=8$. The 7 K model identifies this observed behavior with a
crossing between the two-fold degenerate A valleys in upper spin
level of the lowest LL ($N = 0$) and the lowest spin level of the
second LL ($N = 1$). This crossing occurs when $E_Z = E_C$.
Assuming the effective mass equals $m_{\mathcal{D}} = 0.358m_{o}$,
we obtain an enhanced $g^* = 3.3$ (bare value $g=2$). When these
parameters are used, the 7 K model can explain the observed
strengthening of the $\nu = 6$ minimum with increasing $\theta$
prior to the level crossing and, although not shown, explains why
$\nu = 12$ is never observed.

     The 7 K model with $m_{\mathcal{D}} = 0.358m_o$, implies $\sim$ 50\%
of the electrons are occupying the lowest two valleys at small $B$
fields. If only these contributed to the low $B$ SdH oscillations,
a periodicity of $\Delta \nu\simeq8$ would be expected if the
degeneracy due to spin and valley pairs with opposite $\mathbf{k}$
remain unresolved. Figure 3c shows the remainder of the
coincidence measurements, where $d^2R_{(12,34)}/dB_{\perp}^2$ is
plotted versus $1/\cos \theta$ (x-axis) and $\nu$ (y-axis). The
plot is a compilation of $R_{(12,34)}$ measurements oriented at 45
different $\theta$'s and evidence of up to $\sim5$ level crossings
is apparent. A schematic energy fan diagram in accordance with
expectations from the 7 K model, $\Delta\nu = 8$ and $g^*= 3.3$,
is overlaid on top of the data, and the intersections of the
levels coincides with minima in $d^2R_{(12,34)}/dB_{\perp}^2$ or
maxima in $R_{xx}$. Figure 3d shows calculations derived from the
7 K model of the derivative of the Fermi energy with respect to
the density, $dE_F/dn$, in the same range plotted in Figure 3c.
The same fan diagram used in Figure 3c is also overlaid on top of
the calculations.  By applying different Dingle temperatures, 0.6
K and 2 K, to the carriers originating from the two-fold and
four-fold valleys respectively, the 7 K model is able to reproduce
the unresolved spin degeneracy and the absence of oscillations
from the higher energy valleys seen in the data. Differing
scattering times are commonly observed in two component electron
systems with different densities\cite{houten}.


    A likely origin for breaking the valley degeneracy in our
2DES is the miscut of the wafer. This miscut angle, $\psi$, can be
derived from the width of the atomic steps on a H-Si(111) surface.
Figure 1b shows an atomic force microscope (AFM) image of a
H-Si(111) surface obtained from an adjacent piece of the Si(111)
wafer used to fabricate the device. From the AFM image we observe
steps running at an angle, $\phi \simeq 15 \pm 5^{\circ}$, off
from the [$11\bar{2}$] axis with an average width of $\sim$ 90 nm.
This corresponds to a miscut of $\psi\sim0.2^{\circ}$ and is
approximately about the [$1\bar{1}0$] axis. In the effective mass
approximation, this miscut increases the normal mass, $m_z$, of
the two valleys (A) aligned with the [$11\bar{2}$] direction by
$\sim 0.5$\% relative to the remaining four valleys (B and C). As
a result, the six-fold symmetry is broken into a 4-2 valley
configuration similar to our 7 K model, but with an energy gap of
only $\Delta_{4-2} \sim 2$ K. Additionally, the four valleys (B
and C) occupied at higher energies are split due to the off axis
rotation of $\phi$. This energy gap is calculated to be much
smaller, $\Delta_{4}\sim 0.2$ K, which is within our experimental
uncertainties, but potentially explains the presence of $\nu = 4$.
It is possible that the discrepancy in the magnitude of $\Delta_2$
is due to a many body enhancement. However, these predicted energy
gaps do not include possible effects the atomic steps at the
surface may have on the confinement of the electronic
wavefunction\cite{friesen}.

\begin{figure}[!]
\begin{center}
  \includegraphics[height=2.4in,width=2.9in]{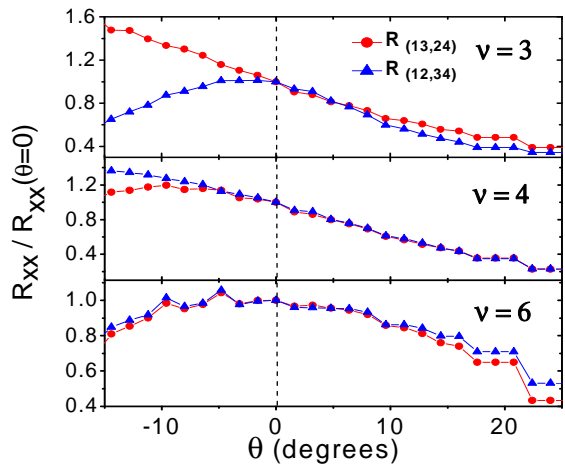}
  \caption{$R_{xx}$ minima at $\nu=3,4,$ and 6 are divided by their
respective $R_{xx}$ values at $\theta=0^{\circ}$ and plotted
within $-15^{\circ}\leq\theta\leq25^{\circ}$. Data represents
measurements at $T = 300$ mK and $n = 6.5 \times 10^{11}$
cm$^{-2}$. } \label{fig4}
\end{center}
\vspace{-26pt}
\end{figure}

    The simple 7 K model does break down at high $B$,
specifically for the observed minima at $\nu = 5, 4,$ and 3.
Figure 4 displays the behavior of the $R_{xx}$ minima for $\nu =
6, 4,$ and 3 versus small $\theta$.  In agreement with the 7 K
model, the $\nu = 6$ minimum strengthens as $\sim 1/\cos \theta$
for small positive and negative tilts about $\theta=0^{\circ}$.
However, the $R_{xx}$ minima for $\nu = 4$ are observed to have a
linear response to small tilts about $B_{\perp}$.  For surface
electrons which have non-zero off diagonal terms normal to the
surface in the effective mass tensor, it is possible to have
confinement energies depending linearly on $\theta$\cite{stern}.
The magnitude of this effect is $\propto \sin \theta \cos \theta$
and has a peak value of $\sim 1.5$ K for $\nu = 4$.  This agrees
well with the observed linear response around small $\theta$ for
$\nu = 4$.

    The observation of $\nu = 3$ and 5 indicates that the in-plane
symmetry of $\pm \,\mathbf{k}_{\parallel}$ between opposite
valleys is broken on our H-Si(111) surface.  We also observe $\nu
= 3$ to be the only state which exhibits anisotropy in $\Delta$
between the two resistance orientations (Table 1). Figure 4 shows
that this anisotropy is also dependent on the measurement
orientation with respect to $B_{\parallel}$.  This splitting
cannot be explained by uniform strain or misorientation effects in
the effective mass approximation. It also cannot be easily
interpreted using theories developed for valley splittings on
[100] Si surfaces\cite{ando,friesen}, where the confinement
potential couples valleys with minima in the direction normal to
the surface. It is possible that disorder on the surface couples
the opposite $\mathbf{k}_{\parallel}$ valleys in our samples.
However, anisotropies of $R_{xx}$ developing at low temperatures
are also observed in other high mobility 2D systems
\cite{lily,pan} and have been attributed to correlated electron
phases. The physics may also be similar to the lifting of the
sublattice symmetry recently observed in graphene\cite{graphene},
or to skyrmion states in multivalley electron
systems\cite{shayegan,arovas}.

    We note that our sample fabrication process is not yet optimized
and that prospects for substantially improved mobilities are good.
Our results imply that the valley splitting on the Si(111)
surfaces can be tuned by wafer orientation and in-plane magnetic
fields.  Thus, H-passivated Si surfaces can be prepared with the
maximum number of valley degrees of freedom in which to explore
potentially new many-body quantum phenomena, such as exotic
fractional quantum Hall states\cite{lai1}, and `valley singlet'
states which are potentially relevant for topological quantum
computing\cite{topo}. Lastly, the relative ease of incorporating
surface engineering within the fabrication process may enable
future experiments in which high mobility electrons are coupled to
molecules or atoms on an atomically clean semiconductor surface.


Funding was provided by the National Security Agency.
\vspace{-15pt}


\end{document}